\documentclass[12pt]{article}
\usepackage{epsfig}
\usepackage{amssymb}
\usepackage{amsmath}
\usepackage{amsfonts}
\usepackage{graphicx}
\usepackage{mathrsfs}
\usepackage[dvips]{color}
\usepackage{multirow}

\newcommand{\R}{\mathbb{R}}
\newcommand{\C}{\mathbb{C}}

\newcommand{\ff}{\mathfrak{f}}
\newcommand{\fg}{\mathfrak{g}}

\newcommand{\fz}{\mathfrak{z}}

\newcommand{\fK}{\mathfrak{K}}

\newcommand{\bfa}{\mathbf{a}}

\newcommand{\bk}{\mathbf{k}}

\newcommand{\bp}{{\mathbf{p}}}

\newcommand{\bx}{\mathbf{x}}

\newcommand{\bA}{\mathbf{A}}

\newcommand{\bB}{\mathbf{B}}

\newcommand{\bI}{\mathbf{I}}

\newcommand{\bM}{\mathbf{M}}

\newcommand{\cH}{\mathcal{H}}
\newcommand{\cE}{\mathcal{E}}

\newcommand{\cG}{\mathcal{G}}

\newcommand{\cK}{\mathcal{K}}
\newcommand{\cL}{\mathcal{L}}

\newcommand{\cP}{\mathcal{P}}

\newcommand{\cT}{\mathcal{T}}

\newcommand{\be}{\begin{equation}}
\newcommand{\ee}{\end{equation}}
\newcommand{\bea}{\begin{eqnarray}}
\newcommand{\eea}{\end{eqnarray}}
\newcommand{\nn}{\nonumber}
\newcommand{\kt}{\rangle}
\newcommand{\br}{\langle}

\newcommand{\ed}{\end{document}}

\newcommand{\bi}{\begin{itemize}}
\newcommand{\ei}{\end{itemize}}

\newcommand{\bce}{\begin{center}}
\newcommand{\ece}{\end{center}}

\newcommand{\sE}{\mathscr{E}}
\newcommand{\sF}{\mathscr{F}}

\newcommand{\sH}{\mathscr{H}}

\newcommand{\sV}{\mathscr{V}}

\newcommand{\bcK}{{\boldsymbol{\cK}}}
\newcommand{\bcE}{{\boldsymbol{\cE}}}
\newcommand{\bcG}{{\boldsymbol{\cG}}}

\newcommand{\bPhi}{{\boldsymbol{\Phi}}}
\newcommand{\bPsi}{{\boldsymbol{\Psi}}}

\newcommand{\for}{{\rm for}}

\begin{document}

\title{Scattering by a collection of $\delta$-function point and parallel line defects in two dimensions}

\author{Hai Viet Bui\thanks{E-mail address: haibui@utexas.edu}~, 
Farhang Loran\thanks{E-mail address: loran@iut.ac.ir}~, and
Ali~Mostafazadeh\thanks{E-mail address:
amostafazadeh@ku.edu.tr}\\[6pt]
$^{*}$Department of Chemistry and Physics, Augusta University,\\
1120 15th Street, Augusta, GA 30912, US\\[6pt]
$^\dagger$Department of Physics, Isfahan University of Technology, \\ Isfahan 84156-83111, Iran\\[6pt]
$^\ddagger$Departments of Mathematics and Physics, Ko\c{c}
University,\\  34450 Sar{\i}yer, Istanbul, Turkey}

\date{ }
\maketitle

\begin{abstract}

Interaction of waves with point and line defects are usually described by $\delta$-function potentials supported on points or lines. In two dimensions, the scattering problem for a finite collection of point defects or parallel line defects is exactly solvable. This is not true when both point and parallel line defects are present. We offer a detailed treatment of the scattering problem for finite collections of point and parallel line defects in two dimensions. In particular, we perform the necessary renormalization of the coupling constants of the point defects, introduce an approximation scheme which allows for an analytic calculation of the scattering amplitude and Green's function for the corresponding singular potential, investigate the consequences of perturbing this potential, and comment on the application of our results in the study of the geometric scattering of a particle moving on a curved surface containing point and line defects. Our results provide a basic framework for the study of spectral singularities and the corresponding lasing and antilasing phenomena in two-dimensional optical systems involving lossy and/or active thin wires and parallel thin plates.




\end{abstract}

\section{Introduction}
\label{S1}

Wave propagation and scattering in effectively two-dimensional media that contain point and line defects is of great interest in condensed matter physics \cite{kosevich-2000,steigerwald-2009,yao-2009,basant-2020} and photonics \cite{sugimoto-2004,sugitatsu-2004,noda-2006,serebryannikov-2008}. Analytic treatment of these phenomenon usually involves idealized models where the interaction of the wave with the defects are described by $\delta$-function potentials supported on a discrete set of points or lines. For the cases where the number of defects are finite and the line defects are parallel, we can choose a Cartesian coordinate system in which the interaction potential takes the form, $V=V_1+V_2$, where
	\begin{align}
	&V_1(\bx):=\sum_{n=1}^N\zeta_n\delta(\bx-\bfa_n),
	&&V_2(\bx):=\sum_{j=1}^J\xi_j\delta(x-b_j),
	\end{align}
$\bx:=(x,y)$ is the position vector, $N$ and $J$ are respectively the numbers of the point and line defects, $\bx=\bfa_n$ and $x=b_j$ mark their position, and $\zeta_n$ and $\xi_j$ are real or complex coupling constants.\footnote{We also assume that the point defects do not lie on the line defects, i.e., $b_j$ is different from the $x$-component of $\bfa_n$ for all $j\in\{1,2,\cdots,J\}$ and all $n\in\{1,2,\cdots,N\}$.}

Unlike its one-dimensional analog, the study of the spectral and scattering properties of the multi-delta-function potential $V_1$ meets serious difficulties. Specifically, in trying to solve the corresponding Lippmann-Schwinger equation one encounters divergent quantities. The basic reason for the emergence of these unwanted divergences is that this potential does not define a self-adjoint Hamiltonian operator. A remedy is offered by von~Neumann's theory of self-adjoint extensions of symmetric operators \cite{Bonneau-2001,albaverio}. Alternatively, one can adopt an appropriate renormalization scheme to remove the divergent terms \cite{mead-1991,manuel,Adhikari1,Adhikari2,Henderson,Mitra,Nyeo, Camblong,teo,ap-2019}. This leads to an exact closed-form expression for the scattering amplitude of $V_1$, \cite{ap-2019}. Solving the scattering problem for the potential $V_2$ is much easier; one can reduce it to a one-dimensional problem and derive an exact formula for its scattering amplitude \cite{epjp-2021}. 

A rather surprising and less known fact is that unlike the potentials $V_1$ and $V_2$, the standard treatment of the scattering problem for their sum, $V_1+V_2$, does not lead to an exact solution.\footnote{This is even true for $N=J=1$.} In this article we offer a comprehensive treatment of this problem. In particular, we outline a non-perturbative approximation scheme for its solution which is valid whenever the distance between the point defect(s) to the nearest line defect(s) is much larger than the wavelength of the incident wave. We use this scheme to determine the purely out-going Green's function for the potential $V_1+V_2$ which in turn allows us to use the first Born approximation to determine the effect of adding a small perturbation to $V_1+V_2$. 

The present investigation is motivated by our interest in the study of the consequences of the presence of point and line defects on the geometric scattering of a particle moving on a curved surface \cite{pra-1996,pra-2018}. Refs.~\cite{ap-2019,epjp-2021} consider the geometric scattering problem for the cases that the surface includes either point defects or a collection of parallel line defects. Following the approach of \cite{ap-2019,epjp-2021}, we can determine the geometric scattering amplitude for a surface containing both point and parallel line defects by identifying the contribution of its geometry with a perturbation $\delta V$ of the potential $V_1+V_2$.

The organization of this articles is as follows. In Sec.~\ref{S2} we present some general properties of the relevant resolvent operators which yield the Green's function for the out-going waves. In Sec.~\ref{S3}, we offer a self-contained treatment of the scattering problem for the potential $V_1$, where we discuss the regularization of the divergences and a corresponding coupling-constant renormalization, and obtain the scattering amplitude of this potential.  In Sec.~\ref{S4}, we treat the scattering problem for $V_1+V_2$ and devise an approximation scheme for its solution. In Sec.~\ref{S5}, we explore the consequences of adding a small perturbation to $V_1+V_2$. Sec.~\ref{S6} includes our concluding remarks. In the appendices we give the details of a technical calculation and develop an extension of the approximation scheme of Sec.~4 which is capable of computing higher-order corrections to its outcome.

\section{Resolvent operators and Lippmann-Schwinger equation}
\label{S2}

Consider a quantum system defined by the Hamiltonian operator $\hat H=\hat H_0+\hat V$, where $\hat H_0$ is the free Hamiltonian and $\hat V$ is the interaction potential. Suppose that $\hat H$ admits scattering states of energy $E>0$, and introduce the resolvent operators,
	\begin{align}
	&\hat G_0:=\lim_{\epsilon\to 0^+}\frac{1}{E-\hat H_0+i\epsilon},
	&&\hat G:=\lim_{\epsilon\to 0^+}\frac{1}{E-\hat H+i\epsilon}.
	\label{e1}
	\end{align}
Then, according to the latter relation,
	\bea
	1+\hat V \hat G&=&\lim_{\epsilon\to 0^+} \left[ (E-\hat H_0+i\epsilon)
	\left(\frac{1}{E-\hat H+i\epsilon}\right)\right],
	\label{e2.a}\\
	1+\hat G\hat V&=&\lim_{\epsilon\to 0^+} \left[ 
	\left(\frac{1}{E-\hat H+i\epsilon}\right)(E-\hat H_0+i\epsilon)\right].
	\label{e2.b}
	\eea
Applying $\hat G_0$ to both sides of (\ref{e2.a}) from the left and to both sides of (\ref{e2.b}) from the right, we arrive at 
	\bea
	\hat G&=&\hat G_0+\hat G_0\hat V\hat G,
	\label{id1}\\
	\hat G&=&\hat G_0+\hat G\,\hat V \hat G_0.
	\label{id1b}
	\eea
Among the useful implications of these relations are the Born series for $\hat G$,	
	\be
	\hat G=\hat G_0\sum_{\ell=0}^\infty (\hat V\hat G_0)^\ell,
	\label{Born}
	\ee
which we can obtain by repeated use of (\ref{id1}), and the identity
	\be
	(1+\hat G\hat V)(1-\hat G_0\hat V)=1,
	\label{id1c}
	\ee
which follows from (\ref{id1b}).	
	
The scattering solutions of the time-dependent Schr\"odinger equation, $\hat H|\psi\kt=E|\psi\kt$, satisfy the Lippmann-Schwinger equation,
	\be
	|\psi\kt=|\psi_0\kt+\hat G_0\hat V|\psi\kt,
	\label{LS-eq}
	\ee
where $|\psi_0\kt$ is a solution of the time-dependent Schr\"odinger equation, $\hat H_0|\psi\kt=E|\psi\kt$. With the help of (\ref{id1c}), we can write the solution of this equation in the form
	\be
	|\psi\kt=|\psi_0\kt+\hat G\hat V|\psi_0\kt.
	\label{LS-sol}
	\ee
Here $|\psi_0\kt$ and $\hat G\hat V|\psi_0\kt$ respectively correspond to the incident and scattered waves. Substituting (\ref{Born}) in (\ref{LS-sol}), we find the Born series for the scattering solution. The $N$-th order Born approximation corresponds to neglecting all but the first $N+1$ terms of this series.  For example, the first Born approximation yields,
	\be
	|\psi\kt=|\psi_0\kt+\hat G_0\hat V|\psi_0\kt.
	\label{1st-Born}
	\ee

The above discussion extends to situations where $\hat H$ and $\hat H_0$ are Hamiltonian operators with the same continuous spectrum. In particular, $\hat H_0$ may consist of a free (kinetic energy) part and an interaction potential.

\section{Scattering by point defects in two dimensions}
\label{S3}
	
For a scalar particle that moves in the $x$-$y$ plane and interacts with $N$ point defects located at $\bfa_1, \bfa_2,\cdots,\bfa_N$, the free Hamiltonian and the interaction potential respectively read
	\begin{align}
	&\hat H_0:=\frac{\hat\bp^2}{2m},
	&\hat V=\hat V_1:=\sum_{n=1}^N\zeta_n|\bfa_n\kt\br\bfa_n|.
	\label{H-pt}
	\end{align}
Here $\hat\bp$ is the standard momentum operator acting in the space of square-integrable functions $L^2(\R^2)$, and $\zeta_n$ are real or complex coupling constants \cite{ap-2019}. 

In the position representation, the Schr\"odinger equation, $\hat H|\psi\kt=E|\psi\kt$, takes the form $L_1\psi(\bx)=0$, where
	\begin{align}
	&L_1:=L_0-\sum_{n=1}^N\fz_n\delta(\bx-\bfa_n),
	&&L_0:=\nabla^2+k^2,\\
	&k:=\frac{\sqrt{2m E}}{\hbar},
	&&\fz_n:=\frac{2m\zeta_n}{\hbar^2}.
	\end{align}
This follows from,
	\begin{align}
	&\br\bx|(\hat H_0-E)=-\frac{\hbar^2}{2m}\:L_0\br\bx|,
	&&\br\bx|(\hat H-E)=-\frac{\hbar^2}{2m}\:L_1\br\bx|.
	\nn
	\end{align} 
If we use $\hat G_1$ to denote the resolvent operator $\hat G$ for $\hat V=\hat V_1$, we can identify 
	\begin{align}
	&\cG_0(\bx,\bx'):=\frac{\hbar^2}{2m}\,\br\bx|\hat G_0|\bx'\kt,
	&&\cG_1(\bx,\bx'):=\frac{\hbar^2}{2m}\,\br\bx|\hat G_1|\bx'\kt,
	\label{e2}
	\end{align}
respectively with the Green's functions associated with the differential operators $L_0$ and $L_1$; they fulfill
	\be
	L_0\,\cG_0(\bx,\bx')=L_1\,\cG_1(\bx,\bx')=\delta(\bx-\bx').\nn
	\ee
	
Suppose that $\bk$ is the wave vector for the incident wave, so that $|\psi_0\kt=|\bk\kt$. Then in view of (\ref{H-pt}), the position wave function for the scattering state vector (\ref{LS-sol}) takes the form,
	\bea
	\br\bx|\psi\kt=\br\bx|\psi_1\kt&:=&
	\br\bx|\bk\kt+\sum_{n=1}^N\zeta_n
	\br\bx|\hat G_1|\bfa_n\kt\br\bfa_n|\bk\kt\nn\\
	&=&\frac{1}{2\pi}\left[e^{i\bk\cdot\bx}+
	\sum_{n=1}^N\fz_n\cG_1(\bx,\bfa_n)\,
	e^{i\bfa_n\cdot\bk}\right].
	\label{scatt-sln}
	\eea
This relation reduces the solution of the scattering problem for the multi-delta function potential to the determination of the Green's function $\cG_1(\bx,\bx')$.
	
To calculate $\cG_1(\bx,\bx')$, first we recall the following well-known consequence of (\ref{e1}), (\ref{H-pt}), and (\ref{e2}).
	\be
	\cG_0(\bx,\bx')=-\frac{i}{4}\,H^{(1)}_0(k\left|\bx-\bx'\right|),
	\label{e3}
	\ee
where $H^{(1)}_0$ stands for the zero-order Hankel function of the first kind. With the help of (\ref{H-pt}) and (\ref{e2}), we can express (\ref{id1}) in terms of the Green's functions $\cG_0$ and $\cG_1$. This gives
    \be
    \cG_1 (\bx,\bx')=\cG_0(\bx,\bx')+\sum_{n=1}^N\fz_n \,
    \cG_0(\bx,\bfa_n)\, \cG_1 (\bfa_n,\bx').
    \label{e4}
    \ee
Setting $\bx=\bfa_m$ in this equation, we find the following linear system of equations for $X_n(\bx'):=\fz_n\cG_1(\bfa_n,\bx')$.
	\be
	\sum_{n=1}^NA_{nm}X_n(\bx')=\cG_0(\bfa_m,\bx'),
	\label{e5}
	\ee
where 
	\begin{align}
	&A_{nm}:=\fz_n^{-1}\delta_{mn}-\cG_0(\bfa_m,\bfa_n)
	=\fz_n^{-1}\delta_{mn}+\frac{i}{4}\,
	H_0^{(1)}(k|\bfa_m-\bfa_n|).
	\label{e6}
	\end{align}
A major difficulty in dealing with (\ref{e5}) is that, because $H^{(1)}_0(0)=\infty$, the coefficients $A_{nm}$ blow up for $m=n$. This calls for a regularization of these coefficients and a renormalization of the coupling constants $\fz_n$. 

Following the approach of Ref.~\cite{ap-2019}, we suppose that the coupling constants $\fz_n$ depend on a real and positive running parameter $\rho$, which represents the size of the point defects, set $x=\bfa_m+\rho$ in (\ref{e4}), and explore the small-$\rho$ asymptotics of the resulting equations. This gives (\ref{e5}) with
	\be
	A_{mn}:=\frac{1}{4}\left\{\begin{array}{ccc}
	4\tilde\fz_n^{-1}+i & \for & m=n,\\[6pt]
	i H_0^{(1)}(k|\bfa_m-\bfa_n|)& \for & m\neq n,
	\end{array}\right.
	\label{Mnm=}
	\ee
where $\tilde\fz_n$ are the renormalized coupling constants given by,
	\be
	\tilde\fz_n:=\frac{2\pi\fz_n}{2\pi-\fz_n[\ln(k\rho/2)+\gamma]},\nn
	\ee
$\gamma$ is the Euler number, and we have exploited the asymptotic expression,
	\be
	H^{(1)}_0(x)=\frac{2i[\ln(x/2)+\gamma]}{\pi}+1+O(x^2).\nn
	\ee
	
Eq.~(\ref{e5}) has a unique solution provided that the determinant of the matrix $\bA$ of its coefficients $A_{mn}$ is nonzero. This corresponds to the values of the wavenumber $k$ where the Hamiltonian $\hat H$ has no spectral singularities \cite{prl-2009}, i.e., scattering amplitude is non-singular. In this case, $\bA$ is an invertible matrix, and we can express the solution of (\ref{e5}) in the form
	\be
	X_n(\bx')=\sum_{m=1}^{N}A^{-1}_{nm}\,\cG_0(\bfa_m,\bx'),
	\label{X=}
	\ee
where $A_{nm}^{-1}$ denote the entries of $\bA^{-1}$. Notice that because $X_n(\bx'):=\fz_n\cG_1(\bfa_n,\bx')$, we can use (\ref{X=}) to express (\ref{e4}) as
	\be
	\cG_1 (\bx,\bx')=\cG_0(\bx,\bx')+
    	\sum_{n,m=1}^N\cG_0(\bx,\bfa_n)\, A^{-1}_{nm}\,
	\cG_0(\bx',\bfa_m).
    \label{e4n}
    \ee
The fact that $\bA$ and consequetly $\bA^{-1}$ are symmetric matrices implies that the right-hand side of (\ref{e4n}) is invariant under an exchange of $\bx$ and $\bx'$. Hence, 
	\be
	\cG_1 (\bx,\bx')=\cG_1 (\bx',\bx),
	\label{G1-sym}
	\ee
and $\fz_n\cG_1(\bx,\bfa_n)=X_n(\bx)$. With the help of the latter relation and (\ref{X=}), we can write (\ref{scatt-sln}) in the form
	\bea
	\br\bx|\psi_1\kt&=&
	\frac{1}{2\pi}\left[e^{i\bk\cdot\bx}+
	\sum_{m,n=1}^N e^{i\bfa_n\cdot\bk} A^{-1}_{nm}\,
	\cG_0(\bx,\bfa_m)\right].
	\label{scatt-sln=}
	\eea	
	
Next, we recall that according to (\ref{LS-sol}) the scattered wave is given by $\psi_s(\bx):=\br\bx|\hat G_1\hat V|\bk\kt$, and that in two dimensions the scattering amplitude $\ff(\bk',\bk)$ along the wave vector $\bk':=k\bx/|\bx|$ satisfies
	\be
	\psi_s(\bx)\to\frac{\ff(\bk',\bk)\,e^{ikr}}{2\pi \sqrt r}~~~\for~~~r\to\infty,
	\label{scatt-ampl-def}
	\ee
where $r:=|\bx|$. Noting that $\psi_s(\bx)$ corresponds to the second term on the right-hand side of (\ref{scatt-sln=}) and making use of the asymptotic formula, 
	\be
	H^{(1)}_0(k|\bx-\bx'|)\to\sqrt{\frac{2}{\pi i k r}}\: e^{-i\bk'\cdot\bx'}
	e^{ikr}~~~\for~~~r\to\infty,
	\label{asym-Ha}
	\ee
which in light of (\ref{e3}) implies
	\be
	\cG_0(\bx,\bx')\to-\sqrt{\frac{i}{8\pi k r}}\: e^{-i\bk'\cdot\bx'}
	e^{ikr}~~~\for~~~r\to\infty,
	\label{asym-H}
	\ee
we arrive at
	\be
	\ff(\bk',\bk)=\ff_1(\bk',\bk):=-\sqrt{\frac{i}{8\pi k}}
	\sum_{m,n=1}^N A^{-1}_{nm}e^{i(\bk\cdot\bfa_n-\bk'\cdot\bfa_m)}.
	\label{f1=}
	\ee
This relation coincides with Eq.~(32) of Ref.~\cite{ap-2019}.\footnote{Ref.~\cite{pra-2016} proposes an alternative treatment of a single-delta-function potential in two dimensions which avoids the singularities of the standard approach and reproduces (\ref{f1=}) for $N=1$. Ref.~\cite{jpa-2018} extends the results of Ref.~\cite{pra-2016} to potentials consisting of a linear array of $\delta$ functions in two-dimensions.}

\section{Scattering by a collection of point and line defects}
\label{S4}

Suppose that the scalar particle we considered in the preceding section also interacts with a finite number of parallel line defects. Assuming that the latter lie along the lines given by $x=b_j$ for some $b_1,b_2,\cdots,b_J\in\R$, we can express the interaction potential in the form
	\be
	\hat V:=\hat V_1+\hat V_2,
	\label{V-total}
	\ee
where $\hat V_1$ is given by (\ref{H-pt}),
	\begin{align}
	&\hat V_2:=\sum_{j=1}^J\xi_j\int_{-\infty}^\infty\!\!dy\: |b_j,y\kt\br b_j,y|=
	\sum_{j=1}^J\xi_j |b_j\kt\br b_j|\otimes \hat I_2,
	\label{V2=}
	\end{align}
$\xi_j$ are real or complex coupling constants, $|x,y\kt$ is the position ket $|\bx\kt$ for $\bx:=(x,y)$, i.e., $|x,y\kt:=|x\kt\otimes|y\kt$, and $\hat I_2$ is the identity operator acting on the space of square-integrable functions of $y$, i.e.,
$\hat I_2:=\int_{-\infty}^\infty\!dy\,|y\kt\br y|$. Clearly,
	\begin{align}
	&\br\bx|\hat V_1=\sum_{n=1}^N\zeta_n\delta(\bx-\bfa_n)\br\bx|,
	&&\br\bx|\hat V_2=\sum_{j=1}^J\xi_j\delta(x-b_j)\br\bx|.\nn
	\end{align}
	
Let us introduce $\hat H_1:=\hat H_0+\hat V_1$, so that the Hamiltonian operator takes the form,
	\be
	\hat H=\hat H_1+\hat V_2.
	\label{H=12}
	\ee
To compute the resolvent operator $\hat G$ for this Hamiltonian, we pursue the approach of the preceding section with $\hat H_1$, $\hat V_2$, and $\hat G_1$ playing the role of $\hat H_0$, $\hat V_1$, and $\hat G_0$, respectively. Then, in view of (\ref{id1}) and (\ref{LS-sol}), the scattering state vectors $|\psi_2\kt$ for the Hamiltonian (\ref{H=12}) satisfy
	\bea
	|\psi_2\kt&=&|\psi_1\kt+\hat G_1\hat V_2|\psi_2\kt
	\label{e101}\\
	&=&|\psi_1\kt+\hat G\hat V_2|\psi_1\kt.
	\label{e102}
	\eea
It is easy to see that,
	\be
	\cG(\bx,\bx'):=\frac{\hbar^2}{2m}\br\bx|\hat G|\bx'\kt,
	\label{e104}
	\ee
gives the Green's function associated with out-going solutions of $L_2\psi_2(\bx)=0$, where 
	\be
	L_2:=L_1-\sum_{j=1}^J\fg_j\delta(x-b_j)=
	\nabla^2+k^2-\sum_{n=1}^N\fz_n\delta(\bx-\bfa_n)-
	\sum_{j=1}^J\fg_j\delta(x-b_j),
	\label{e103}
	\ee
and $\fg_j:={2m}\xi_j/\hbar^2$. 

To determine the scattering solutions of the Schr\"odinger equation, $H|\psi_2\kt=E|\psi_2\kt$, we examine the position representation of (\ref{e101}), i.e.,
	\be
	\psi_2(x,y)=\psi_1(x,y)+\sum_{j=1}^J\fg_j\int_{-\infty}^\infty \!dy'\:
	\cG_1(x,y;b_j,y')\psi_2(b_j,y').
	\label{e105}
	\ee
Here we have employed (\ref{V2=}), expressed $\bx$ and $\bx'$ in terms of their components $(x,y)$ and $(x',y')$, and used $\cG_1(x,y;x',y')$ to abbreviate $\cG_1((x,y),(x',y'))$. Setting $x=b_i$ in (\ref{e105}), we obtain
the following system of integral equations for $\psi_2(b_j,y)$.
	\be
	\psi_2(b_i,y)-\sum_{j=1}^J \fg_j\int_{-\infty}^\infty\!\! dy' \left[
	 \cG_1(b_i,y;b_j,y')\psi_2(b_j,y')\right]=\psi_1(b_i,y).
	\label{e105b}
	\ee
In view of (\ref{e4n}),  we can express the $\cG_1(b_i,y;b_j,y')$ appearing in this relation in the form,
	\bea
	\cG_1(b_i,y;b_j,y')&=&\cG_0(b_i,y;b_j,y')+\delta\cG_{ij}(y,y'),
	\label{e106}
    	\eea
where 
	\begin{align}
	&\cG_0(x,y;x',y'):=\cG_0\big((x,y),(x',y')\big)=
	-\frac{i}{4}H^{(1)}_0\Big(\mbox{$k\sqrt{(x-x')^2+(y-y')^2}$}\Big),
	\label{e107}\\
	&\delta\cG_{ij}(y,y'):=\sum_{n,m=1}^N 
	\cG_0 (b_i,y;a_{nx},a_{ny})
	\,A^{-1}_{nm}\,
	\cG_0 (b_j,y';a_{mx},a_{my}),
	\label{e107b}
	\end{align}	
and $a_{nx}$ and $a_{ny}$ are respectively the $x$- and $y$-components of $\bfa_n$.

The integral equations (\ref{e105b}) do not admit a closed-form analytic solution. In the following, we obtain an approximate solution of these equations that applies to situations where the distance between the nearest point and line defects is much larger than the wavelength of the incident wave.

Let $\ell$ be the minimum of the set,
	\[\Big\{|a_{nx}-b_j|~\Big|~n\in\{1,2,\cdots,N\}, j\in\{1,2,\cdots,J\}\Big\},\]
and suppose that $k\ell\gg 1$. Then, according to (\ref{asym-H}) and (\ref{e107}), $\Big|\cG_0 (b_i,y;a_{nx},a_{ny})\,
	\cG_0 (b_j,y';a_{mx},a_{my})\Big|$ is bounded by a multiple of $1/k\ell$;	
	\be
	\Big|\cG_0 (b_i,y;a_{nx},a_{ny})\,
	\cG_0 (b_j,y';a_{mx},a_{my})\Big|=
	O\left(\mbox{\large$\frac{1}{k\ell}$}\right).
	\label{approx-o}
	\ee
This allows us to neglect $\delta\cG_{ij}(y,y')$ in (\ref{e106}), and conclude that
	\be
	\cG_1(b_i,y;b_j,y')\approx \cG_0(b_i,y;b_j,y')~~~\for~~~k\ell\gg 1.
	\label{approx}
	\ee
As we explain below, this approximation reduces (\ref{e105b}) to a system of algebraic equations. This follows from the identity, 
	\be
	\cG_0(x,y;x',y')=-\frac{i}{4\pi}\int_{-\infty}^\infty\!\!
	d\fK\left[\frac{e^{i\sqrt{k^2-\fK^2}|x-x'|}
	\, e^{i\fK(y-y')}}{\sqrt{k^2-\fK^2}}\right],
	\label{id-app}
	\ee
whose derivation we give in Appendix~A. Using (\ref{id-app}) together with the definition of the Fourier and inverse Fourier transform in one dimension, i.e.,
\begin{align}
	&\sF_{y,\fK}\{f(y)\}:=\int_{-\infty}^\infty dy\: e^{-i\fK y}f(y),
	&&\sF^{-1}_{\fK,y}\{\tilde f(\fK)\}:=\frac{1}{2\pi}
	\int_{-\infty}^\infty d\fK\: e^{i\fK y}\tilde f(\fK),
	\label{Fourier}
	\end{align}
we find
	\be
	\int_{-\infty}^\infty\!\!dy' \cG_0(b_i,y;b_j,y')\psi_2(b_j,y')=
	\sF^{-1}_{\fK,y}\big\{\cE(b_i-b_j,\fK)\, Y_j(\fK)\big\},
	\label{e109}
	\ee
where 
	\bea
	\cE(x,\fK)&:=&-\frac{i\,\displaystyle e^{i\sqrt{k^2-\fK^2}\,|x|}}{2\sqrt{k^2-\fK^2}},
	\label{cE-def}
	\\[6pt]
	Y_j(\fK)&:=&\sF_{y,\fK}\left\{\psi_2(b_j,y)\right\}.
	\label{Yj}
	\eea 
	
In view of (\ref{e109}), if we take the Fourier transform of both sides of (\ref{e105b}), we obtain
	\be
	\sum_{j=1}^J B_{ij}(\fK)Y_j(\fK)\approx\tilde\psi_1(b_i,\fK),
	\label{e110}
	\ee
where
	\begin{align}
	 &B_{ij}(\fK):=\delta_{ij}-\fg_j\, \cE(b_i-b_j,\fK),
	&&\tilde\psi_1(b_i,\fK):=\sF_{y,\fK}\left\{\psi_1(b_i,y)\right\}.
	\label{e111}
	\end{align}
Substituting (\ref{scatt-sln=}) in the latter relation and using (\ref{Fourier}) and (\ref{id-app}), we arrive at the following more explicit expression for $\tilde\psi_1(b_i,\fK)$.
	\be
	\tilde\psi_1(b_i,\fK)=e^{ib_ik_x}\delta(\fK-k_y)
	+\frac{1}{2\pi}\sum_{m,n=1}^N
	e^{i\bfa_n\cdot\bk}A^{-1}_{nm}\,e^{-i a_{my}\fK}\,
	\cE(a_{mx}-b_i,\fK).
	\label{e111b}
	\ee
	
According to (\ref{e110}), 
	\be
	Y_i(\fK)\approx\sum_{j=1}^J B_{ij}^{-1}(\fK)\tilde\psi_1(b_j,\fK),
	\label{Yi=}
	\ee
where $ B_{ij}^{-1}(\fK)$ are the entries of the inverse of the matrix $\bB(\fK):=[B_{ij}(\fK)]$. In light of (\ref{Yj}) and (\ref{Yi=}), 
	\be
	\psi_2(b_i,y)\approx\sum_{j=1}^J
	\sF^{-1}_{\fK,y}\left\{ B_{ij}^{-1}(\fK)\tilde\psi_1(b_j,\fK)
	\right\}=
	\sum_{j=1}^J\int_{-\infty}^\infty\!\! dy' 
	\check B_{ij}(y-y')\psi_1(b_j,y'),
	\label{e112}
	\ee
where 
	\be \check B_{ij}(y):=\sF_{\fK,y}^{-1}
	\left\{B_{ij}^{-1}(\fK)\right\},
	\label{check-B}
	\ee
and we have made use of the convolution formula for the Fourier transform. 

Next, we introduce
	\be
	\tilde\cG_1(\bx;b_i,\fK):=\sF_{y',\fK}\{\cG_1(x,y;b_i,y')\},
	\label{tcG1-def}
	\ee
and use (\ref{Fourier}) and (\ref{e112}) to show that
	\bea
	\int_{-\infty}^\infty\!\!dy'\cG_1(x,y;b_i,y')\psi_2(b_i,y')
	&\approx&
	\sum_{j=1}^J\int_{-\infty}^\infty\!\!dy'\,
	\sF_{\fK',y'}^{-1}\{\tilde\cG_1(\bx;b_i,\fK')\}
	\sF^{-1}_{\fK,y'}\left\{ B_{ij}^{-1}(\fK)\tilde\psi_1(b_j,\fK)
	\right\}\nn\\
	&\approx&\frac{1}{2\pi}\sum_{j=1}^J\int_{-\infty}^\infty\!\!d\fK\:
	\tilde\cG_1(\bx;b_i,-\fK)B_{ij}^{-1}(\fK)\tilde\psi_1(b_j,\fK).
	\eea
In view of this relation and (\ref{e105}), 
	\be
	\psi_2(x,y)\approx\psi_1(x,y)+\frac{1}{2\pi}\sum_{i,j=1}^J
	\int_{-\infty}^\infty\!\!d\fK\:
	\tilde\cG_1(\bx;b_i,-\fK)B_{ij}^{-1}(\fK)\tilde\psi_1(b_j,\fK).
	\label{e105c}
	\ee
To arrive at a more explicit expression for the right-hand side of this relation, we note that according to (\ref{id-app}), (\ref{Fourier}), and (\ref{cE-def}),
	\be
	\sF_{y,\fK}\left\{\cG_0(x,y;x',y')\right\}
	=e^{-i\fK y'}\cE(x-x',\fK),
	\label{id-tG0}
	\ee
and use (\ref{e4n}), (\ref{tcG1-def}), (\ref{id-tG0}), and $\cG_0(\bx,\bx')=\cG_0(\bx',\bx)$ to establish
	\bea
	\tilde\cG_1(\bx;b_i,\fK)
	&=&e^{-i\fK y}\cE(b_i-x,\fK)+
    	\sum_{n,m=1}^N\cG_0(x,y;a_{nx},a_{n,y})\, A^{-1}_{nm}
	e^{-i a_{my}\fK}\,\cE(b_i-a_{mx},\fK).
	\label{tcG1=}
	\eea
Substituting (\ref{e111b}) and (\ref{tcG1=}) in (\ref{e105c}), we can determine the scattering solution $\psi_2(x,y)$ for $k\ell\gg 1$.  

Next, we use (\ref{scatt-sln=}) and the expression (\ref{e105}) for $\psi(x,y)$ to find the scattered wave $\psi_s(x,y)$ and use its asymptotic expression (\ref{scatt-ampl-def}) to calculate the scattering amplitude. This yields
	\be
	\ff(\bk',\bk)= \ff_1(\bk',\bk)+\ff_2(\bk',\bk),
	\label{f=f+f}
	\ee
where $\ff_1(\bk',\bk)$ is given by (\ref{f1=}), 
	\bea
	\ff_2(\bk',\bk)&:=&-\sqrt{\frac{i\pi}{2k}}
	\sum_{j=1}^J\fg_j\left[e^{-ik'_x b_j}Y_j(k'_y)
	+\sum_{m,n=1}^N 
	e^{-i \bk'\cdot\bfa_n}A^{-1}_{nm}C_{jm}
	\right],
	\label{f2=}\\
	C_{jm}&:=&\int_{-\infty}^\infty \!\!dy'\:
	\cG_0(a_{mx},a_{my};b_j,y')\psi_2(b_j,y')
	=\sF_{\fK,a_{my}}^{-1}
	\big\{\cE(a_{mx}-b_j,\fK)Y_j(\fK)\big\},
	\label{Cjm}
	\eea	
and we have used (\ref{e4n}), (\ref{asym-H}), (\ref{e105}), (\ref{Fourier}), and (\ref{e109}). 

Eq.~(\ref{f=f+f}) with $\ff_1(\bk',\bk)$, $\ff_2(\bk',\bk)$, and $C_{jm}$ given by (\ref{f1=}), (\ref{f2=}) and (\ref{Cjm}) provide an exact expression for the scattering amplitude of the potential (\ref{V-total}). These relations are however of little practical value unless we are also able to compute $Y_j(\fK)$. 
For cases where $k\ell\gg 1$, we can use (\ref{Yi=}) to obtain an approximate formula for the latter. Let us recall that to derive this formula, we have neglected terms bounded by constant multiples of $1/k\ell$. According to (\ref{asym-Ha}) and (\ref{e107}), whenever $k\ell\gg 1$, $|\cG_0(a_{mx},a_{my};b_j,y')|$ is bounded by terms proportional to $1/\sqrt{k\ell}$. Eq.~(\ref{Cjm}) implies that the same applies to the coefficients $C_{jm}$. If $k\ell$ is so large that we can also neglect $C_{jm}$ the scattering amplitude for the potential (\ref{V-total}) reduces to the sum of the scattering amplitudes for the potentials $V_1$ and $V_2$. This is however not true, if $1/k\ell$ is negligible but $1/\sqrt{k\ell}$ is not. 

The large-$k\ell$ approximation we have introduced above may be extended to an approximation scheme which allows for computing corrections of higher order in powers of $1/k\ell$. We outline this scheme in Appendix~B.

\section{Perturbing the potential for point and line defects}
\label{S5}

Consider a perturbation of the potential (\ref{V-total});
	\be
	\hat V\to \hat\sV:=\hat V_1+\hat V_2+\delta\hat V,
	\label{e201}
	\ee
where $\hat V_1$ and $\hat V_2$ are respectively given by (\ref{H-pt}) and (\ref{V-total}), and suppose that $\delta\hat V$ is such that the first-order Born approximation provides a reliable description of the scattering problem for $\hat\sV$. 

Let $\hat\sH:=\hat H+\delta\hat V$, where
	\begin{align}
	&\hat H:=\frac{\hat\bp^2}{2m}+\hat V_1+\hat V_2,
	\label{H=2}
	\end{align}
and $\hat G$ be the unperturbed resolvent operator given by (\ref{e1}) for the Hamiltonian~(\ref{H=2}). Then, in view of (\ref{1st-Born}) and the fact that $|\psi_2\kt$ of Eq.~(\ref{e105}) are scattering solutions for the Schr\"odinger equation, $\hat H|\psi_2\kt=E|\psi_2\kt$, the first-order Born approximation yields the following expression for the scattering solutions of the Schr\"odinger equation, $\hat\sH|\psi\kt=E|\psi\kt$.
	\be
	|\psi\kt=|\psi_2\kt+\hat G\delta \hat V|\psi_2\kt.
	\label{1st-Born-2}
	\ee

Now, recall that the scattering amplitude $\ff(\bk',\bk)$ for the perturbed potential satisfies (\ref{scatt-ampl-def}) and that the scattered wave is defined by: $\br\bx|\psi_s\kt=\br\bx|\psi\kt-\br\bx|\bk\kt$. These observations together with Eq.~(\ref{1st-Born-2}) suggest that
	\be
	\ff(\bk',\bk)=\ff_1(\bk',\bk)+\ff_2(\bk',\bk)+\delta\ff(\bk',\bk),
	\label{e202}
	\ee
where $\ff_1(\bk',\bk)$ and $\ff_2(\bk',\bk)$ are respectively given by 
(\ref{f1=}) and (\ref{f2=}), $\delta\ff(\bk',\bk)$ fulfills
	\bea
	\delta\ff(\bk',\bk)&=&\lim_{r\to\infty}\left[
	2\pi\sqrt r\: e^{-ikr}\br\bx|\hat G\delta\hat V|\psi_2\kt\right]\nn\\
	&=&\lim_{r\to\infty}\left[
	\frac{4\pi m\sqrt r\: e^{-ikr}}{\hbar^2}
	\int_{\R^2}d^2\bx'\: \cG(\bx,\bx')\br\bx'|\delta\hat V|\psi_2\kt\right],
	\label{e203}
	\eea
and $\cG(\bx,\bx')$ is the Green's function defined by (\ref{e104}). 

Eqs.~(\ref{e202}) and (\ref{e203}) reduce the determination of the scattering amplitude $\ff(\bk',\bk)$ for the potential (\ref{e201}) to that of $\cG(\bx,\bx')$. 
To achieve the latter, we first recall that according to (\ref{H=12}), $\hat H=\hat H_1+\hat V_2$. This suggests relating $\hat G$ to the resolvent operator $\hat G_1$ for the Hamiltonian $\hat H_1$ by letting $\hat V_2$ and $\hat G_1$ play the role of $\hat V$ and $\hat G_0$ in (\ref{id1}). This implies
	\be
	\hat G=\hat G_1+\hat G_1\hat V_2\hat G.
	\label{id1-2}
	\ee
Equivalently, we have
	\be
	\cG(\bx,\bx')=\cG_1(\bx,\bx')+\int_{\R^2}d^2\tilde\bx\: \cG_1(\bx,\tilde\bx)
	\br\tilde\bx|\hat V_2\hat G|\bx'\kt.
	\label{e204}
	\ee
	
Next, we label the $x$- and $y$-components of $\tilde\bx$ respectively by $\tilde x$ and $\tilde y$, and use (\ref{V2=}) to express (\ref{e204}) as 
	\be
	\cG(x,y;x',y')=\sum_{j=1}^J\fg_j\int_{-\infty}^\infty\!\! 
	d\tilde y\:\cG_1(x,y;b_j,\tilde y)\cG(b_j,\tilde y;x',y')+
	\cG_1(x,y;x',y').
	\label{e205a}
	\ee
For $x=b_i$, this gives
	\be
	\cG(b_i,y;x',y')-\sum_{j=1}^J\fg_j\int_{-\infty}^\infty\!\! 
	d\tilde y\:\cG_1(b_i,y;b_j,\tilde y)\cG(b_j,\tilde y;x',y')=
	\cG_1(b_i,y;x',y').
	\label{e205}
	\ee
When the point defects are at such a large distance from the line defects that $k\ell\gg 1$, we can use the approximation (\ref{approx}) to replace the $\cG_1(b_i,y;b_j,\tilde y)$ in (\ref{e205}) with $\cG_0(b_i,y;b_j,\tilde y)$. Using (\ref{id-app}) and (\ref{Fourier}), we can then express (\ref{e205}) in the form,
	\be
	\cG(b_i,y;x',y')-\frac{1}{2\pi}\sum_{j=1}^J\fg_j
	\int_{-\infty}^\infty\!\! 
	d\fK\: e^{i\fK y}\,\cE(b_i-b_j,\fK)\,Z_j(\fK,\bx')
	\approx\cG_1(b_i,y;x',y'),
	\label{e206}
	\ee
where
	\be
	Z_j(\fK,\bx'):=\sF_{y,\fK}\left\{\cG(b_j,y;x',y')\right\}.
	\label{Zj-def}
	\ee
Taking the one-dimensional Fourier transform of both sides of (\ref{e206}) gives rise to the following system of linear equations for $Z_j(\fK,\bx')$.
	\be
	\sum_{j=1}^J B_{ij}(\fK)Z_j(\fK,\bx')\approx\tilde\cG_1(\bx';b_i,\fK),
	\label{e207}
	\ee
where $B_{ij}(\fK)$ and $\tilde\cG_1(\bx';b_i,\fK)$ are respectively given by (\ref{e111}) and (\ref{tcG1=}), and we have employed (\ref{G1-sym}).
	
The system of equations given by (\ref{e207}) has a unique solution whenever the matrix $\bB(\fK)$ of its coefficients $B_{ij}(\fK)$ has a nonzero determinant. This happens when there are no spectral singularities. In this case,
	\be
	Z_i(\fK,\bx')\approx
	\sum_{j=1}^J B_{ij}^{-1}(\fK)\tilde\cG_1(\bx';b_i,\fK).
	\label{e207b}
	\ee
This relation together with (\ref{check-B}) and (\ref{Zj-def}) imply
	\be
	\cG(b_i,y;x',y')\approx\sum_{j=1}^J
	\sF_{\fK, y}^{-1}\left\{B_{ij}^{-1}(\fK)\tilde\cG_1(\bx';b_i,\fK)\right\}=\sum_{j=1}^J
	\int_{-\infty}^\infty\!\!d\tilde y\:\check B_{ij}(y-\tilde y)\cG_1(b_j,\tilde y;x',y').
	\label{e208}
	\ee
Substituting (\ref{e4n}) and (\ref{e208}) in (\ref{e205a}), we obtain the Green's function $\cG(\bx,\bx')$. This in turn gives
	\bea
	\lim_{r\to\infty}\sqrt r\: e^{-ikr}\cG(\bx,\bx')&\approx&
	-\sqrt{\frac{i}{8\pi k}}
	\left\{e^{-i\bk'\cdot\bx'}+\sum_{m,n=1}^N
	e^{-i\bfa_n\cdot\bk'}A_{nm}^{-1}
	\cG_0(\bx',\bfa_m)+\right.\nn\\
	&&\hspace{2cm}
	\left.\sum_{j=1}^J\fg_j\Big[e^{-ib_jk'_x}Z_j(k'_y,\bx')
	+\sum_{m,n=1}^N e^{-i\bfa_n\cdot\bk'}
	A_{nm}^{-1}D_{mj}\Big]\right\},
	\label{cG-asymp}
	\eea
where we have employed (\ref{e4}) and (\ref{asym-H}), and introduced
	\bea
	D_{mj}&:=&
	\int_{-\infty}^\infty\!\!d\tilde y\:
	\cG_0(a_{mx},a_{my};b_j,\tilde y)\cG(b_j,\tilde y;x',y')
	=\sF_{\fK,a_{my}}^{-1}
	\big\{\cE(a_{mx}-b_j,\fK)\, Z_j(\fK,\bx')\big\}.
	\label{Dmj}
	\eea
Inserting  (\ref{cG-asymp}) in (\ref{e203}) and using (\ref{tcG1=}) and (\ref{e207b}), we can calculate the contribution of the perturbation $\delta\hat V$ to the scattering amplitude for the potential (\ref{e201}), namely
	\bea
	\delta\ff(\bk',\bk)&\approx&
	-\frac{m}{\hbar^2}\sqrt{\frac{2\pi i }{k}} 
	\int_{\R^2}d^2\bx'\: 
	\left\{e^{-i\bk'\cdot\bx'}+\sum_{m,n=1}^N
	e^{-i\bfa_n\cdot\bk'}A_{nm}^{-1}
	\cG_0(\bx',\bfa_m)+\right.\nn\\
	&&\left.\sum_{j=1}^J\fg_j\Big[e^{-ib_jk'_x}Z_j(k'_y,\bx')
	+\sum_{m,n=1}^N e^{-i\bfa_n\cdot\bk'}
	A_{nm}^{-1}D_{mj}\Big]\right\}
	\br\bx'|\delta\hat V|\psi_2\kt.
	\label{e203x}
	\eea

Eqs.~(\ref{f1=}), (\ref{f2=}), (\ref{e202}), and (\ref{e203x}) provide the solution for the scattering problem of the perturbed potentials of the form (\ref{e201}). This also applies to situations where the perturbation $\delta\hat V$ is function of both the position and momentum operators. An interesting example is the effective geometric potential that describes the contribution of the nontrivial geometry of an asymptotically flat curved surface $S$ on the scattering of scalar particles moving on it. In this case, 
	\begin{equation}
    	\br\bx|\,\delta\hat V:=\frac{\hbar^2}{2m}\,\delta L_{\bx}\,\br \bx|,
    	\label{eq23}
    	\end{equation}
where $\cL_{\bx}$ is the differential operator,
	\begin{equation}
    	\delta L_{\bx}:=\nabla^2-\Delta_S+ 2\left[\lambda_1 K(x)+\lambda_2 M(x)^2\right],
   	\label{eq24}
    	\end{equation}
$\nabla^2$ is the Laplacian in two dimensions, $\Delta_S$, $K$, and $M$ are respectively the Laplace-Beltrami operator, the Gaussian curvature, and the mean curvature of the surface, and $\lambda_1$ and $\lambda_2$ are a pair of real coupling constants whose values depends on the details of the confining forces that keeps the particle on the surface \cite{pra-2018}.

\section{Concluding remarks}
\label{S6}

$\delta$-function potential offers an invaluable tool for teaching quantum mechanics. In one dimension this potential and its multi-center generalizations are among a handful of examples that admit an explicit analytic treatment. At the same time they have ample physical applications in the study of defects. In two dimensions, the $\delta$-function potential provides an extremely simple model whose standard treatment leads to the emergence of divergent terms and calls for their removal via a suitable renormalization scheme. This presents an exceptional opportunity for implementing the basic ideas and methods of renormalization theory within the context of non-relativistic quantum mechanics. The same difficulties appear and can be similarly dealt with for the multi-center $\delta$-function potentials.

The single- and multi-$\delta$-function potentials supported on parallel lines in two dimensions provide another class of exactly solvable models with applications in modeling line defects. Their treatment is more straightforward, for it does not involve dealing with unwanted singularities. 

The exact solvability is a common feature of both the (multi-)$\delta$-function potentials supported on points or parallel lines in two dimensions. Surprisingly this feature is lost once we add members of these two classes of singular potentials to those of the other. The latter correspond to a two-dimensional physical system involving both point and line defects. In the present article, we have provided a comprehensive treatment of the scattering problem for such systems. For this purpose, we have developed an approximation scheme that yields an analytic expression for the scattering amplitude. Our approach provides reliable results whenever the distance between the point defects to the nearest line defect is much larger than the wavelength of the incident wave. We have used the same approximation to determine the Green's function for the $\delta$-function potentials describing collections of point and parallel line defects. This in turn enables us to compute the scattering amplitude for the small perturbations of these potentials. 

The large-separation approximation we have developed is not sensitive to the distance between different point defects or different line defects. Perhaps more importantly, as we show in Appendix~B, it admits a hierarchal generalization that is capable of computing higher-order corrections in the powers of the small parameter of the approximation, namely $1/k\ell$. 

An interesting application of our results is in the study of effectively two-dimensional optical systems involving thin wires and parallel thin plates. In particular, for the cases where the thin wires (respectively thin plates) are made of gain material, the condition that the matrix $\bA$ (respectively $\bB$) be singular corresponds to the emergence of a spectral singularity \cite{prl-2009} and marks the onset of lasing \cite{pra-2011a,ramezani-2014,Moccia-2020,josab-2020}. This happens for situations where some of the coupling constants of the potentials $V_1$ and $V_2$ have a positive imaginary part. The time-reversal of this effect, which corresponds to complex-conjugation of the coupling constants yielding a spectral singularity, is known as coherent perfect absorption or antilasing \cite{longhi-2010-CPA-laser,wan-2011,chong-2011,jpa-2012,kalozoumis-2017,Konotop-2019b}. Our results provide a characterization of spectral singularities of the potential $V_1+V_2$ and open the way for the study of lasing and antilasing in effectively two-dimensional optical systems involving thin wires and parallel thin plates made of active or lossy material. 

Our approach for treating the scattering by point and parallel line defects in two dimensions may be extended to the study of finite collections of line and parallel planar defects in three dimensions. A more difficult task is to devise a similar approach for dealing with collections of point, parallel line, and parallel planar defects. This would require a renormalization of the coupling constants for the $\delta$-function potentials modeling point and line defects.

\section*{Appendix~A: Derivation of (\ref{id-app})}

To derive  (\ref{id-app}), we begin recalling that
	\begin{align}
	&\cG_0(x,y;x',y')=\lim_{\epsilon\to 0}\left[
	\frac{\hbar^2}{2m}\br x,y|\left(E-\frac{\hat\bp^2}{2m}+i\epsilon\right)^{-1}|x',y'\kt\right],
	\label{app-e1}\\
	&|x,y\kt=|x\kt\otimes|y\kt,\quad\quad\quad|x',y'\kt=|x'\kt\otimes|y'\kt.\nn
	\end{align}
We let $\hat k_x:=\hat p_x/\hbar$ and $\hat k_y:=\hat p_y/\hbar$, so that $|k_x,k_y\kt=|k_x\kt\otimes|k_y\kt$, $\br k_y|\hat p_y=\hbar k_y\br k_y|$,
	\begin{align}
	&\br y|k_y\kt=\frac{e^{ik_y y}}{\sqrt{2\pi}}, 
	&\int_{-\infty}^\infty\!\! dk_y\:|k_y\kt\br k_y|=\hat I_2, 
	&&\br k_y|\hat\bp^2=\br k_y|(\hat p_x^2+p_y^2)=\br k_y|
	\left(\hat p_x^2+\hbar^2 k_y^2 \right).\nn
	\end{align}
Using these relation in (\ref{app-e1}), we have
	\bea
	\cG_0(x,y;x',y')&=&
	\lim_{\epsilon\to 0^+}
	\left[\frac{\hbar^2}{2m}\br x|\otimes\int_{-\infty}^\infty\!\! dk_y
	\br y|k_y\kt \br k_y|\left(E-\frac{\hat\bp^2}{2m}+i\epsilon\right)^{-1}
	|y'\kt\right]\otimes|x'\kt\nn\\
	&=&
	\frac{1}{2\pi}\int_{-\infty}^\infty\!\! dk_y\:
	\left[ e^{ik_y(y-y')}\br x|\lim_{\tilde\epsilon\to 0^+}
	\left(k^2-k_y^2-\hat k_x^2+i\tilde\epsilon\right)^{-1}|x'\kt\right].
	\label{app-e2} 
	\eea
We can identify $\lim_{\tilde\epsilon\to 0^+}\left(\sE-\hat k_x^2+i\tilde\epsilon\right)^{-1}$ with the resolvent operator for the Hamiltonian $H:=\hat k_x^2$. In other words $\br x|\lim_{\tilde\epsilon\to 0^+}
	\left(\sE-\hat k_x^2+i\tilde\epsilon\right)^{-1}|x'\kt$ is the Green's function for the differential operator $\frac{d^2}{dx^2}+\sE$ which has the following well-known expression
	\[\br x|\lim_{\tilde\epsilon\to 0^+}
	\left(\sE-\hat k_x^2+i\tilde\epsilon\right)^{-1}|x'\kt=
	-\frac{i e^{i \sqrt\sE |x-x'|}}{2 \sqrt\sE}.\]
Substituting this relation with $\sE=k^2-k_y^2$ in (\ref{app-e2}) and setting $\fK:=k_y$, we arrive at (\ref{id-app}).

\section*{Appendix~B: A perturbative series solution for (\ref{e105b}) and (\ref{e205}) }

In Sec.~\ref{S4}, we have obtained an approximate solution (\ref{e105b}) which involves neglecting $\delta\cG_{ij}(y,y')$ on the right-hand side of (\ref{e106}) for the cases where $k\ell\gg 1$. We have employed the same approximation in Sec.~\ref{S5} while solving (\ref{e205}) for $\cG(b_i,y;x',y')$. In this appendix, we outline formal series solutions of (\ref{e105b}) and (\ref{e205}) whose first terms reproduce the approximate solutions we obtain in Secs.~\ref{S4} and \ref{S5}.

Eqs.~(\ref{e105b}) and (\ref{e205}) are examples of the following system of integral equations.
	\be
	\Psi_i(y)-\int_{-\infty}^\infty\!\!
	dy'\Big\{\sum_{j=1}^J\fg_j\left[\cG_{0ij}(y-y')+\delta
	\cG_{ij}(y,y')\right]\Psi_j(y')\Big\}=\Phi_i(y),
	\label{e401}
	\ee
where $\Phi_i:\R\to\C$ and $\Psi_i:\R\to\C$, with $i\in\{1,2,\cdots,i\}$, are respectively the known and unknown functions, and
	\[\cG_{0ij}(y):=\cG_0(b_i,y;b_j,0)=-\frac{i}{4}H_0^{(1)}\Big(
	k\sqrt{(b_i-b_j)^2+y^2}\Big).\]
Eqs.~(\ref{e401}) give (\ref{e105b}) [respectively (\ref{e205})] provided that we set $\Phi_i(y):=\psi_1(b_i,y)$ and $\Psi_i(y):=\psi_2(b_i,y)$ [respectively $\Phi_i(y):=\cG_1(b_i,y;x',y')$ and $\Psi_i(y):=\cG(b_i,y;x',y')$.]

We begin our analysis by introducing the $J$-component functions
	\begin{align}
	&\bPhi(y):=\left[\begin{array}{c}
	\Phi_1(y)\\
	\Phi_2(y)\\
	\vdots\\
	\Phi_{_{\!J}}(y)\end{array}\right],
	&&\bPsi(y):=\left[\begin{array}{c}
	\Psi_1(y)\\
	\Psi_2(y)\\
	\vdots\\
	\Psi_{_{\!J}}(y)\end{array}\right],
	\end{align}
and $J\times J$ matrix-valued functions 
	\begin{align}
	&\bcG_0(y):=[\cG_{0ij}(y)], &&\delta\bcG(y,y'):=[\delta\cG_{ij}(y,y')].
	\label{bGs-def}
	\end{align} 
Then, letting $\cH$ denote the space of $J$-component functions and adopting Dirac's bra-ket notion, we can write (\ref{e401}) as the following linear equation in $\cH$.
	\be
	(\widehat\bI-\widehat\bcG_0-\widehat{\delta\bcG})|\bPsi\kt=|\bPhi\kt,
	\label{e402}
	\ee
where $\widehat\bI$ is the identity operator, and $\widehat\bcG_0$ and $\widehat{\delta\bcG}$ are operators acting in $\cH$ according to
	\begin{align}
	&\br y|\widehat\bcG_0|\bPsi\kt:=\int_{-\infty}^\infty\!\! dy'\:\bcG_0(y-y')\bPsi(y'),
	&&\br y|\widehat{\delta\bcG}|\bPsi\kt:=\int_{-\infty}^\infty\!\! dy'\:\delta\bcG(y,y')\bPsi(y').
	\label{hbGs-def}
	\end{align}
Supposing that $\widehat\bI-\widehat\bcG_0$ is invertible and expressing its inverse by $\widehat\bcK$, we can write (\ref{e402}) as
	\be
	(\widehat\bI-\widehat\bcK\,
	\widehat{\delta\bcG})|\bPsi\kt=\widehat\bcK|\bPhi\kt.
	\label{e403}
	\ee
This in turn leads us to the following formal series solution of (\ref{e402}).
	\be
	|\bPsi\kt=\sum_{n=0}^\infty \big(\widehat\bcK\,
	\widehat{\delta\bcG}\big)^n\,\widehat\bcK|\bPhi\kt.
	\label{e404}
	\ee

As we describe in Sec.~\ref{S4}, for $k\ell\gg 1$, $|\delta\cG_{ij}(y,y')|$ are bounded by terms proportional to $1/k\ell$. This suggests approximating the series solution (\ref{e404}) by its truncations,
	\be
	|\bPsi\kt\approx|\bPsi^{(N)}\kt:=\sum_{n=0}^N (\widehat\bcK\,
	\widehat{\delta\bcG})^n\widehat\bcK|\bPhi\kt
	=\sum_{n=0}^N (\widehat\bcK\,
	\widehat{\delta\bcG})^n|\bPsi^{(0)}\kt,
	\label{e405}
	\ee
which involve neglecting terms proportional to $1/(k\ell)^{N+1}$. We therefore identify $N$ with the order of the approximation. 

The zeroth-order approximation, 
	\be
	|\bPsi\kt\approx|\bPsi^{(0)}\kt:=\widehat\bcK|\bPhi\kt,
	\label{zero-approx=}
	\ee 
which is equivalent to $(\widehat\bI-\widehat\bcG_0)|\bPsi\kt\approx|\bPhi\kt$, corresponds to (\ref{e402}) [respectively (\ref{e401})] with the term $\widehat{\delta\bcG}|\bPsi\kt$ [respectively $\delta\cG_{ij}(y,y')$] missing. This is precisely the approximation we have employed in trying to solve equations (\ref{e105b}) and (\ref{e205}). Using the approach we pursued to obtain (\ref{Yi=}), we can show that 
	\be
	\widehat\bcK=\int_{-\infty}^\infty\!\! d\fK\:
	\bB(\fK)^{-1}|\fK\kt\br\fK|,
	\label{bcK=1}
	\ee
where $\bB(\fK)$ is the $J\times J$ matrix with entries $B_{ij}(\fK)$ given by (\ref{e111}), and 
	\be
	\br y|\fK\kt:=\frac{e^{i\fK y}}{\sqrt{2\pi}}.
	\label{k-ef}
	\ee
Inserting (\ref{bcK=1}) in (\ref{zero-approx=}), we find
	\be
	|\Psi^{(0)}\kt=
	\int_{-\infty}^\infty\!\! d\fK\:
	\bB(\fK)^{-1}\br\fK|\bPhi\kt\:|\fK\kt.
	\label{Psi-zero=}
	\ee

The fact that, according to (\ref{bcK=1}), $\widehat\bcK$ is diagonal in the $\fK$-representation suggests us to examine the $\fK$-representation of $\widehat{\delta\bcG}$. In view of (\ref{e107b}), (\ref{id-tG0}), (\ref{bGs-def}), and (\ref{k-ef}), we find the following expression for the entries of the matrix $\br\fK|\widehat{\delta\bcG}|\fK'\kt$.
	\bea
	\br\fK|\widehat{\delta\cG}_{ij}|\fK'\kt&=&
	\int_{-\infty}^\infty\!\!dy\:\int_{-\infty}^\infty\!\!dy'
	\br\fK|y\kt\,\delta\cG_{ij}(y,y')\,\br y'|\fK'\kt\nn\\
	&=&\frac{1}{2\pi}\sum_{m,n=1}^N
	\int_{-\infty}^\infty\!\!dy\:\int_{-\infty}^\infty\!\!dy'
	e^{-i(\fK y-\fK' y')}\cG_0(b_i,y;a_{nx},a_{ny})\,A_{nm}^{-1}\,
	\cG_0(b_j,y';a_{mx},a_{my})\nn\\
	&=&\frac{1}{2\pi}\sum_{m,n=1}^N
	\sF_{y,\fK}\left\{\cG_0(b_i,y;a_{nx},a_{ny})\right\}
	A_{nm}^{-1}\,
	\sF_{y',-\fK'}\left\{\cG_0(b_j,y';a_{mx},a_{my})\right\}\nn\\
	&=&\frac{1}{2\pi}\sum_{m,n=1}^N
	e^{-i\fK a_{ny}}\cE(b_i-a_{nx},\fK)\,A_{nm}^{-1}\,
	e^{i\fK' a_{my}}\cE(b_j-a_{mx},-\fK').
	\label{e406}
	\eea
Next, we introduce the $J\times 1$ matrices,
	\[\bcE_n(\fK):=e^{-i\fK a_{ny}}
	\left[\begin{array}{c}
	\cE(b_1-a_{nx},\fK)\\
	\cE(b_2-a_{nx},\fK)\\
	\vdots\\
	\cE(b_{_J}-a_{nx},\fK)\end{array}\right],\]
and use them together with (\ref{e406}) to establish,
	\be
	\br\fK|\widehat{\delta\bcG}|\fK'\kt=
	\frac{1}{2\pi}\sum_{m,n=1}^N
	\bcE_n(\fK)\,A_{nm}^{-1}\,
	\bcE_m(-\fK')^T,
	\label{e407}	
	\ee
where $\bM^T$ denotes the transpose of a matrix $\bM$. Eqs.~(\ref{bcK=1}) and (\ref{e407}) imply
	\be
	\widehat\bcK\,\widehat{\delta\bcG}=
	\frac{1}{2\pi}\int_{-\infty}^\infty\!\! d\fK
	\int_{-\infty}^\infty\!\! d\fK' \sum_{m,n=1}^N
	\bB(\fK)^{-1}\bcE_n(\fK)\,A_{nm}^{-1}\,
	\bcE_m(-\fK')^T |\fK\kt\br\fK'|.
	\label{e409}
	\ee
Substituting (\ref{Psi-zero=}) and (\ref{e409})) in (\ref{e405}), we can determine the $N$-th order approximate solution $|\bPsi^{(N)}\kt$ of (\ref{e402}) for $N\geq 1$. 
\vspace{12pt}

\noindent{\bf Acknowledgements:} We wish to thank Kaan G\"uven for suggesting a few relevant references. This work has been supported by the Scientific  and Technological Research Council of Turkey (T\"UB{$\dot{\rm I}$}TAK) in the framework of the Project
No.~117F108 and by the Turkish Academy of Sciences (T\"UBA).
\vspace{12pt}

    

\ed
\begin{thebibliography}{99}    

\bibitem{kosevich-2000} A.~M.~Kosevich and D.~V.~Matsokin,
Low Temperature Phys.\ {\bf 26}, 449 (2000).

\bibitem{steigerwald-2009} 
A.~Steigerwald et al, Appl.\ Phys.\ Lett.\ {\bf 94}, 111910 (2009).

\bibitem{yao-2009} 
Z.-J.~Yao, G.-L.~Yu,Y.-S.~Wang, and Z.-F.~Shi, Int.~J.~Solids \& Structures {\bf 46} 2571 (2009).

\bibitem{basant-2020}
S.~B.~Lal and M.~Gaurav, Phil.\ Trans.~R.\ Soc.~A, 37820190102 (2020).

\bibitem{sugimoto-2004}
Y.~Sugimoto, Y.~Tanaka, N.~Ikeda, Y.~Nakamura, K.~Asakawa, and K.~Inoue, Opt.\ Express {\bf 12} 1090 (2004)

\bibitem{sugitatsu-2004}
A.~Sugitatsu, T.~Asano, and S.~Noda,
Appl.\ Phys.\ Lett.\ {\bf 84}, 5395 (2004).

\bibitem{noda-2006} S.~Noda, J.~Lightwave Tech.\ {\bf 24}, 4554 (2006).

\bibitem{serebryannikov-2008}
A.~E.~Serebryannikov and T. Magath, J.~Opt.\ Soc.\ Am.~B {\bf 25}, 286 (2008).

\bibitem{Bonneau-2001} G.~Bonneau, J.~Faraut, and G.~Valent, Am.\ J.~Phys.~{\bf 69},  322 (2001).

\bibitem{albaverio} S.~Albeverio, F.~Gesztesy, R.~Hoegh-Krohn, and H.~Holden, Solvable Models in Quantum Mechanics (American Mathematical Society, Providence, RI, 2005).

\bibitem{mead-1991} L.~R.~Mead, and J.~Godines, Am.\ J.~Phys.~{\bf 59}, 935 (1991).

\bibitem{manuel} C.~Manuel and R.~Tarrach, Phys.\ Lett.~B {\bf 328} 113 (1994).

\bibitem{Adhikari1} S.~K.~Adhikari and T.~ Frederico, Phys.\ Rev.\ Lett.~\textbf{74}, 4572 (1995).

\bibitem{Adhikari2} S.~Adhikari, T.~Frederico, and  R.~M.~Marinho,  J. Phys. A~\textbf{29}, 7157 (1996).

\bibitem{Henderson} R.~J.~Henderson and S.~G.~Rajeev, J.~Math.\ Phys.\ \textbf{38}, 2171 (1997).

\bibitem{Mitra} I.~Mitra, A.~DasGupta, and B.~Dutta-Roy, Am. J. Phys. \textbf{66}, 1101 (1998).

\bibitem{Nyeo} S.~Nyeo, Am.\ J.~Phys.~\textbf{68},  571 (2000).

\bibitem{Camblong} H.~E.~Camblong and  C.~R.~Ord\'{o}n\~{e}z, Phys. Rev. A \textbf{65}, 052123 (2002).

\bibitem{teo} F.~Erman and O.~T.~Turgut, J.~Phys.~A {\bf 43}, 335204 (2010).
     
\bibitem{ap-2019} H.~Bui and A.~Mostafazadeh, Ann. Phys. (NY) 407, 228-249 (2019).

\bibitem{epjp-2021} H.~Bui, A.~Mostafazadeh, and S.~Seyman, Eur.~Phys.\ J.~Plus {\bf 136}, 109 (2021).

\bibitem{pra-1996} A. Mostafazadeh, Phys. Rev. A  \textbf{54}, 1165-1170 (1996).

\bibitem{pra-2018} N.~Oflaz, A.~Mostafazadeh, and M.~Ahmady, Phys.\ Rev.\ A \textbf{98}, 022126 (2018).

\bibitem{prl-2009} A.~Mostafazadeh,
Phys.\ Rev.\ Lett.~\textbf{102}, 220402 (2009).

\bibitem{pra-2016} F.~Loran and A.~Mostafazadeh, Phys.\ Rev.~A \textbf{93}, 042707 (2016).

\bibitem{jpa-2018} F.~Loran and A.~Mostafazadeh, 
J.~Phys.~A {\bf 51}, 335302 (2018).

\bibitem{pra-2011a} A.~Mostafazadeh, 
Phys.\ Rev.~A {\bf 83}, 045801 (2011).


\bibitem{ramezani-2014}
H.~Ramezani, H.K. Li, Y.~Wang, X.~Zhang, 
Phys.\ Rev.~Lett.\ \textbf{113}, 263905 (2014).

\bibitem{Moccia-2020} 
M.~Moccia, G.~Castaldi, A.~Alu, and V.~Galdi,
IEEE Trans.\ Antennas and Propagation {\bf 68}, 1704-1716 (2020).

\bibitem{josab-2020} 
H.~Ghaemi-Dizicheh, A.~Mostafazadeh, M.~Sar{\i}saman, 
J.~Opt.\ Soc.\ Am.~{\bf 37}, 2128-2138 (2020).

\bibitem{longhi-2010-CPA-laser} 
S.~Longhi, 
Phys. Rev. A {\bf 82}, 031801 (2010).

\bibitem{wan-2011} 
W.~Wan, Y.~Chong, L.~Ge, H.~Noh, A.~D.~Stone, and H.~Cao, 
Science 331, 889 (2011).

\bibitem{chong-2011} 
Y.~D.~Chong, L.~Ge, and A.~D.~Stone, 
``$\cP\cT$-symmetry breaking and laser-absorber modes in optical scattering systems,''
Phys.\ Rev.\ Lett.\ {\bf 106}, 093902 (2011).

\bibitem{jpa-2012}
A.~Mostafazadeh, 
J.~Phys.\ A: Math.\ Theor.\ {\bf 45}, 444024 (2012).

\bibitem{kalozoumis-2017}
P.~Kalozoumis, C.~Morfonios, G.~Kodaxis, F.~Diakonos, P.~Schmelcher, 
``Emitter and absorber assembly for multiple self-dual operation and directional transparency," 
Appl.\ Phys.\ Lett.\ \textbf{110}, 121106 (2017).

\bibitem{Konotop-2019b} 
V.~V.\ Konotop, E.~Lakshtanov, and B.~Vainberg, 
Phys.\ Rev.~A {\bf 99}, 043838 (2019).

 
\end{thebibliography}
